\documentclass[11pt,preprint]{aastex}
\def\simgt{\hbox{\,\rlap{\raise 0.425ex\hbox{$>$}}\lower 0.65ex\hbox{$\sim$}\,}}
\def\simlt{\hbox{\,\rlap{\raise 0.425ex\hbox{$<$}}\lower 0.65ex\hbox{$\sim$}\,}}

\begin{document}

\title{The Fundamental Surface of Quad Lenses}

       \author{Addishiwot G. Woldesenbet and Liliya L.R. Williams}
       \affil{School pf Physics and Astronomy\\
              University of Minnesota\\
              116 Church Street SE\\
              Minneapolis, MN 55455}
       \email{woldesenbet@physics.umn.edu, llrw@astro.umn.edu}

\begin{abstract}
In a quadruply imaged lens system the angular distribution of images around the lens center is completely described by three relative angles. We show empirically that in the three dimensional space of these angles, spanning $180^\circ\times 180^\circ\times 90^\circ$,  quads from simple two-fold symmetric lenses of arbitrary radial density profile and arbitrary radially dependent ellipticity or external shear define a nearly invariant two dimensional surface.  We give a fitting formula for the surface using SIS+elliptical lensing potential. Various circularly symmetric mass distributions with shear up to $\gamma\sim 0.4$ deviate from it by typically, rms$~\sim 0.1^\circ$, while elliptical mass distributions with ellipticity of up to $e\sim 0.4$ deviate from it by rms$~\sim 1.5^\circ$.  The existence of a near invariant surface gives a new insight into the lensing theory and provides a framework for studying quads. It also allows one to gain information about the lens mass distribution from the image positions alone, without any recourse to mass modeling. As an illustration, we show that about 3/4 of observed galaxy-lens quads do not belong to this surface within observational error, and so require additional external shear or substructure to be modeled adequately.  
\end{abstract}

\section{Introduction}

Recovering projected mass distribution of galaxies and clusters given the images of 
lensed background sources is an important problem, and much effort has been devoted 
to lens mass modeling over the last couple of decades. In this paper we show that useful 
information about the lensing object can be obtained without any recourse to mass modeling. 
We work exclusively with quadruply imaged lens systems (the fifth central image is usually
not detected and is not part of the analysis), and more specifically, with the angular 
distribution of the four point-like images around the lens center, that was introduced in 
\cite{wffb08}. We do not consider image fluxes. 

\begin{figure}
\epsscale{0.95}
\plotone{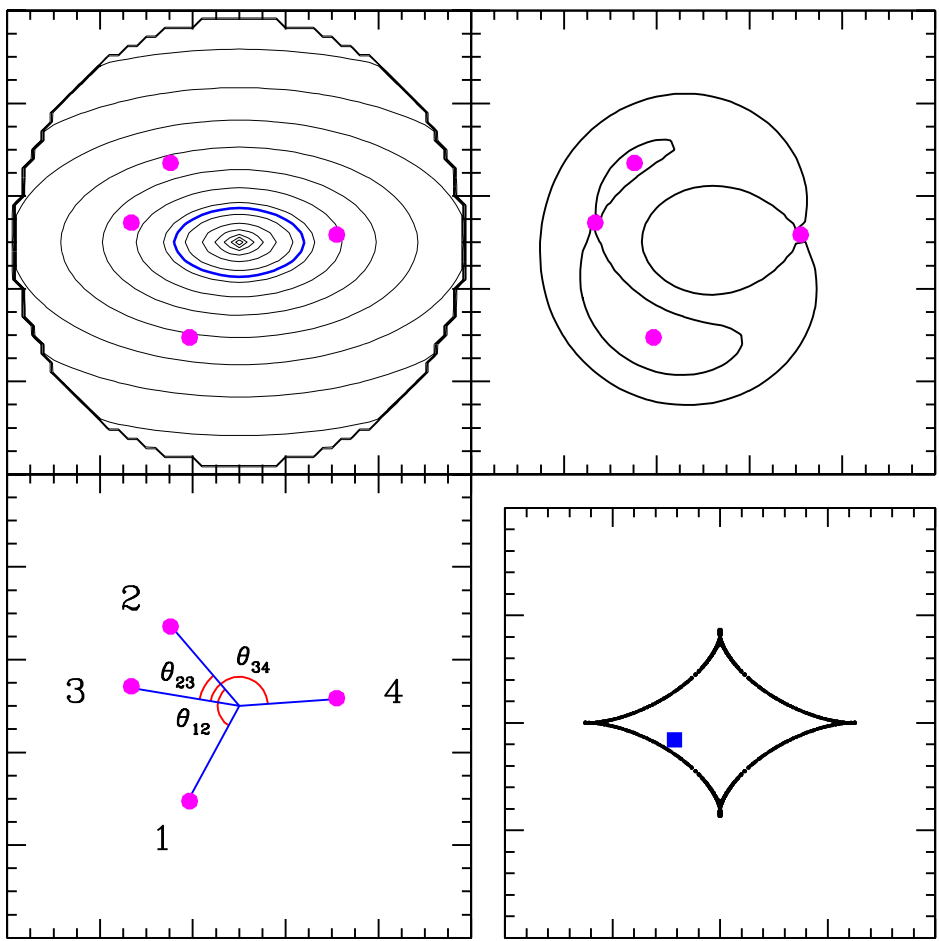}
\vskip-1.3in
\caption{A representative quad from a twofold symmetric lens. 
{\it Top left:} An elliptical lens mass distribution, with the $\kappa=1$ contour
shown as a thick blue line. Images are the magenta filled circles.
{\it Top right:} Arrival time contours and images.
{\it Bottom left:} Images, labeled by arrival time. The relative angles,
$\theta_{12}$, $\theta_{23}$ and $\theta_{34}$ are marked.
{\it Bottom right:} The diamond caustic, with the location of the source 
represented by a solid blue square.}
\label{fig4panels}
\end{figure}

A typical quad image configuration is shown in Figure~\ref{fig4panels}.
The images are labeled by their arrival time at the observer, 1 through 4. 
In most cases this ordering can be determined from the morphology of the quad, 
without measured time delays \citep{sw03}. Here we are interested only in the mass
distribution of the lens, not its total mass (or, equivalently, its normalization),
or orientation. In this case the image configuration of any quad is described 
uniquely by 6 parameters, which we chose to be of the polar variety, measured with 
respect to the lens center: three relative angles, and three distance ratios of images. 
The three relative angles between images are marked on the plot, $\theta_{12}$, 
$\theta_{34}$ and $\theta_{23}$.  Angle $\theta_{12}$ is between the two minima 
of the arrival time surface, while $\theta_{34}$  is the angle between the saddle points.
We define $\theta_{12}$ such that it encloses image 3, and $\theta_{34}$ such that it 
encloses image 2. 
Of the three angles, $\theta_{23}$ is special because the separation between images 2 
and 3 gets arbitrarily small for sources approaching the diamond caustic shown in the lower
right panels of Figure~\ref{fig4panels}; when the source
crosses the caustic these two images disappear and the quad becomes a double.  Note that 
any linearly independent combination of the above three angles can be used, but we chose
$\theta_{12}$, $\theta_{34}$ and $\theta_{23}$ because they have a simple physical meaning.

Working with only two angles, $\theta_{23}$ and a certain linear combination of $\theta_{12}$ 
and $\theta_{34}$, \cite{wffb08} showed that a wide range of simple, twofold symmetric lens 
models generate apparently indistinguishable patterns in the two dimensional plane of these 
angles. Twofold symmetric means that the mass distribution, and hence the potential, is 
symmetric about two orthogonal axes, and 'simple' excludes lenses with 'wavy' isodensity 
shapes. The simple, twofold symmetric class of lenses includes all popular parametric lens 
models, such as Singular Isothermal Ellipsoids (SIE), and Singular Isothermal Elliptical 
Potential (SIEP), as well lenses of any density profile and ellipticity.

The present paper is an extension of \cite{wffb08}, but here we work with the full set of 
three angles, $\theta_{12}$, $\theta_{34}$ and $\theta_{23}$. We show that quads from all 
simple lens mass distributions with twofold symmetry lie on nearly the same two dimensional 
surface in the three dimensional space of relative angles. We call this the Fundamental 
Surface of Quads (FSQ). The quads from observed galaxy lenses, on the other hand, show a
different behavior. As we show in Section~\ref{realq}, galaxy quads form a `cloud' surrounding
the FSQ, with typical separations from the FSQ of few to several degrees. 

One can draw some interesting parallels between the FSQ we introduce here and the well 
studied Fundamental Plane of Ellipticals. Both lie in the three dimensional space whose 
axes are parameters describing the structural properties of the respective objects. 
In the case of quad lenses, these are the relative image angles, while in the case of 
ellipticals they are the effective radius, the surface brightness at the effective radius, 
and the central velocity dispersion. A wide class of objects belong to the Surface and the 
Plane with small scatter. In other words, the objects do not fill the full three dimensional 
space, implying that there is a tight relation between the three parameters. The existence 
of the Fundamental Plane is basically the consequence of the virial theorem, while the 
reason for the near invariance of the Fundamental Surface of Quads for a wide class of 
twofold symmetric lenses (but not necessarily for the observed quads) is yet to be identified.

\section{The SIS+elliptical lensing model}\label{sisell}

We start by studying a simple, analytically tractable, two dimensional projected
gravitational potential. It belongs to the generic family of separable
potentials, $\phi(r,\theta)=r\!\cdot\!f(\theta)$, where $r$ and $\theta$ are
polar coordinates in the lens plane. Properties of such potentials are discussed 
in \citet{Kass}, \cite{k91} and \citet{dalal98}. For our purpose we choose \begin{equation}
f(\theta)=b[1+\gamma\cos(2\theta)]\end{equation}
hence, \begin{equation}
\phi=rb[1+\gamma\cos(2\theta)]\label{SISellpot}\end{equation}
which is sometimes called SIS+elliptical; we will call it SISell for short.
The normalization factor $b$ is the Einstein radius, and $\gamma$ is the shear
parameter.

The Poisson equation, \ensuremath{\Delta}$\phi=2$$\kappa$, yields the projected 
dimensionless mass density profile, 
\begin{equation}
\kappa=\frac{b}{2r}[1-3\gamma\cos(2\theta)].
\end{equation}
Note that $\gamma$ cannot be greater than $1/3$ since otherwise $\kappa$
will have an unphysical negative value. The lens equation, 
\begin{equation}
\vec{r_{s}}=\vec{r}-\vec{\triangledown}\phi\label{eq:a}
\end{equation}
where $\vec{r_{s}}$ and $\vec{r}$ are source and image positions
respectively, can be rewritten as two independent equations 
\begin{equation}
r_{s}\cos(\theta-\theta_{s})=r-b[1+\gamma\cos(2\theta)]\end{equation}
\begin{equation}
r_{s}\sin(\theta-\theta_{s})=-2b\gamma\sin(2\theta)\,\,\,\Rightarrow\,\,\,
-a\sin(\theta-\theta_{s})=\sin(2\theta),\,\,\,\,\,\,{a=\frac{r_{s}}{2b\gamma}}.
\label{eq:lensEquation}\end{equation}

Setting the magnification $M=1/\det(A)$ , where $A$ is the Jacobian matrix
of the lens equation, to infinity, i.e. $\det(A)=0$, one gets
\begin{equation}
\det(A)=\frac{1}{r}\left[\left(r-b\{1+\gamma\cos(2\theta)\}\right)+4b\gamma\cos\left(2\theta\right)\right]=0,
\end{equation}
which yields the condition for the caustic, 
\begin{equation}
r=b-3b\gamma\cos(2\theta).\label{eq:cauticCondition}
\end{equation}
Now using equation (\ref{eq:cauticCondition}) in the lens equation, eq.~\ref{eq:lensEquation},
allows one to express the caustic coordinates $(r_{sc},\theta_{sc})$ in the source plane as a
function of parameter $\theta$,
\begin{eqnarray}
r_{sc} & = & b\gamma\sqrt{2[5+3\cos(4\theta)]}\label{eq:caustic}\\
\theta_{sc} & = & -\tan^{-1}\left(\tan^{3}\theta\right).\label{eq:caustictheta}
\end{eqnarray}

Our aim is to calculate the three relative angles, $\theta_{12}$, $\theta_{34}$ and $\theta_{23}$ 
for all the quads within the diamond caustic. Using eq.~\ref{eq:lensEquation}, which is independent 
of image distance $r$, we get the angular positions of the four images, in radians.
\begin{equation}
\theta_{1}=\cos^{-1}\left(\frac{1}{2}\sqrt{C+\frac{-a^{3}\cos^{3}(\theta_{s})+a\left(a^{2}-4\right)\cos(\theta_{s})+8a\cos(\theta_{s})}{4B}}
-\frac{1}{4}a\cos(\theta_{s})+\frac{B}{2}\right)\label{eqangles1},\end{equation}
 \begin{equation}
\theta_{2}=\cos^{-1}\left(-\frac{1}{2}\sqrt{C-\frac{-a^{3}\cos^{3}(\theta_{s})+a\left(a^{2}-4\right)\cos(\theta_{s})+8a\cos(\theta_{s})}{4B}
}-\frac{1}{4}a\cos(\theta_{s})-\frac{B}{2}\right)\label{eqangles2},\end{equation}
 \begin{equation}
\theta_{3}=\cos^{-1}\left(\frac{1}{2}\sqrt{C-\frac{-a^{3}\cos^{3}(\theta_{s})+a\left(a^{2}-4\right)\cos(\theta_{s})+8a\cos(\theta_{s})}{4B}}
-\frac{1}{4}a\cos(\theta_{s})-\frac{B}{2}\right)\label{eqangles3},\end{equation}
 \begin{equation}
\theta_{4}=-\cos^{-1}\left(-\frac{1}{2}\sqrt{C+\frac{-a^{3}\cos^{3}(\theta_{s})+a\left(a^{2}-4\right)\cos(\theta_{s})+8a\cos(\theta_{s})}{4B
}}-\frac{1}{4}a\cos(\theta_{s})+\frac{B}{2}\right)\label{eqangles4},\end{equation}
where, \begin{equation}
A=\sqrt[3]{108a^{4}\sin^{2}(2\theta_{s})+2\left(a^{2}-4\right)^{3}+12\sqrt{3}\sqrt{a^{4}\sin^{2}(2\theta_{s})\left(27a^{4}\sin^{2}(2\theta_{
s})+\left(a^{2}-4\right)^{3}\right)}},\end{equation}
 \begin{equation}
B=\sqrt{\frac{\left(a^{2}-4\right)^{2}}{62^{2/3}A}+\frac{1}{4}a^{2}\cos^{2}(\theta_{s})+\frac{1}{6}\left(4-a^{2}\right)+\frac{A}{12\sqrt[3]{
2}}},\end{equation}
 \begin{equation}
C=-\frac{\left(a^{2}-4\right)^{2}}{62^{2/3}A}+\frac{1}{2}a^{2}\cos^{2}(\theta_{s})+\frac{1}{3}\left(4-a^{2}\right)-\frac{A}{12\sqrt[3]{2}}.
\end{equation}

The above equations are rather cumbersome and so preclude simple
analytical expressions for the relative angles. Instead, we numerically
generate random source positions within the caustic, calculate $\theta_{i}$'s
using the above equations, and then compute relative angles.

The symmetry of the potential implies that it is sufficient to consider
source positions only within one of the quadrants of the elliptical
potential. Therefore, without any loss of generality we align the
$x$-axis with the major axis of the ellipse and consider $\theta_{s}$
only from $0$ to $\pi/2$. For a given $\theta_{s}$, $r_{s}$ can vary in the range 
$[0$,$r_{sc}$$)$, therefore with eqs.~\ref{eq:lensEquation} and \ref{eq:caustic} we
see that $a$ runs in the range of $\Bigl[0,0.5\sqrt{2[5+3\cos(4\theta)]}\Bigl)$,
which is independent of $b$ and $\gamma$, where the parameter $\theta$
is determined by the value of $\theta_{s}$ using eq.~\ref{eq:caustictheta}.

\section{The Fundamental Surface of Quads}\label{fit}

\begin{figure}
\includegraphics[scale=0.6]{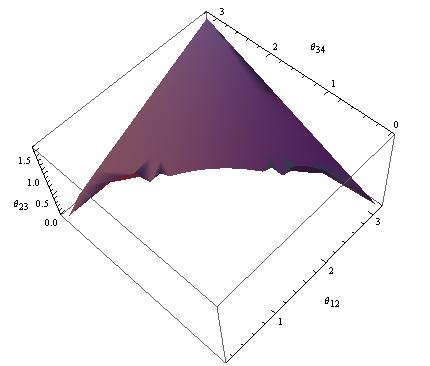}\includegraphics[scale=0.6]{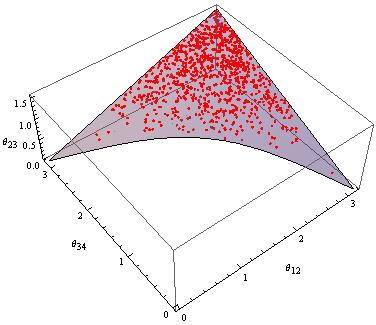}
\caption{The three dimensional space of three image angles, $\theta_{12}$, $\theta_{23}$ and $\theta_{34}$
for the SIS+elliptical, or SISell mass distribution. (a) The surface outlined by quads whose relative image 
angles were calculated using eq.~\ref{eqangles1}-\ref{eqangles4}.  
(b) The fit surface to the Fundamental Surface of Quads (FSQ), 
eq.~\ref{eq:explicitFunction}, is shown as the gray surface, 
while the points are quads randomly distributed within the diamond caustic in the source plane.}
\label{figFPQparam}
\end{figure}

We use the expressions for $\theta_{i}$'s given in the previous Section to parametrically
plot the relative quad angles in the three dimensional space of 
$\theta_{12}$, $\theta_{34}$ and $\theta_{23}$. The resulting distribution is a two 
dimensional surface, shown in Figure~\ref{figFPQparam}(a). The fact that it is a surface 
means that in a quad resulting from a SISell lens, two relative image angles completely 
determine the third. 

The surface is simple with distinct properties. It has a slightly curved 
triangular shape with its apex at ($\theta_{12}$, $\theta_{34}$,
$\theta_{23}$) = ($180^{\circ},180^{\circ},90^{\circ}$); quads at the apex have a {}``cross'' 
configuration. The two edges connecting the apex to the base at $\theta_{23}=0^{\circ}$
correspond to $\theta_{12}=180^{\circ}$ and $\theta_{34}=180^{\circ}$, because in a twofold 
symmetric lens the latter two angles do not exceed $180^{\circ}$. Note that the base of the
triangular surface that represents small values of $\theta_{23}$ and source positions close to 
the caustic, shows some unevenness, or jaggedness. It is unclear whether this is intrinsic to 
$\theta_i$ equations, or if it is due to the numerical noise arising from the implementation of 
these equations. In all what follows we ignore these small features; in particular, our fit to 
the surface, discussed below, smooths over this unevenness.

This surface is universal for all SISell lenses, however the meaning of universality
requires some clarification. For a given source position $(\theta_{s},r_{s})$, a relative 
angle $\theta_{ij}$ depends on $b$ and $\gamma$, and therefore two SISell models with 
different $b$ and $\gamma$ give rise to two different points on the surface. However, the 
surface itself does not depend on $b$ and $\gamma$. This is the result of the elimination 
of $b$ and $\gamma$ dependence when considering all source positions within the caustic as 
discussed in Section \ref{sisell}. Therefore quads from SISell models of all shears and 
normalizations lie on the same invariant surface. 

We would like to have an explicit functional form for the surface,
as $\theta_{23}={\rm fcn}(\theta_{12},\theta_{34})$, but since the equations for individual 
angles, eq.~\ref{eqangles1}-\ref{eqangles4}, contain inverse cosines and are complicated, there 
is no simple expression.  Instead, we calculate thousands of sets of relative angles, 
($\theta_{12}$, $\theta_{34}$, $\theta_{23}$), from the expressions for the $\theta_{i}$'s
(eq.~\ref{eqangles1}-\ref{eqangles4}) and fit these 
with a surface represented by a polynomial function in $\theta_{12}$ and $\theta_{34}$. 
The fitting was done using Matlab's Least Absolute Errors (LAE) method.
As compared to the Least Squares method, LAE is less stable and could generate more than one 
function. We chose LAE anyway because it is resistant to outliers, which in our case
correspond to quads with small values of $\theta_{23}$, and are responsible for the jaggedness 
of the surface. 

We determine the optimal order of the polynomial by considering the deviations, or errors, of the 
SISell quad points from the fit surface, quantified by the root mean square error, RMSE. 
A number of trials and tests revealed that a fourth order polynomial\footnote{Matlab allows up 
to 5th degree polynomial fit.} has the lowest value of RMSE, $\approx0.00032$ radians, or 
$\approx0.018^\circ$, for approximately 160 thousand quads. 
We note that using different sets of quads to obtain the fit equation resulted in slightly
different values for the coefficients of the polynomial, and for some sets of quads the RMSE 
varied by up to a factor of two. 

To test the robustness of our fit to changes in the fitting method, 
we also computed the best fit surface using the Least Square method\footnote{Numerical Recipes'
Singular Value Decomposition routine did not provide a good fit when single precision was 
used, while in double precision \texttt{{svdcmp}} failed to invert the matrix at all.},
and using custom versus in-build polynomials within Matlab. Again, the resulting fit surface
changed somewhat, but did not deviate substantially from the fit we present below. We conclude,
therefore, that although the fit equation is not reproducible exactly, it is completely adequate 
for our purposes:
\begin{eqnarray}
\theta_{23} & = & -5.792+1.783\,\theta_{12}+0.1648\,\theta_{12}^{2}-0.04591\,\theta_{12}^{3}
-0.0001486\,\theta_{12}^{4}+1.784\,\theta_{34}\nonumber \\
 &  & -0.7275\,\theta_{34}\,\theta_{12}+0.0549\,\theta_{34}\,\theta_{12}^{2}
+0.01487\,\theta_{34}\,\theta_{12}^{3}+0.1643\,\theta_{34}^{2}+0.05493\,\theta_{34}^{2}\,\theta_{12}
\label{eq:explicitFunction}\\
 &  & -0.03429\,\theta_{34}^{2}\,\theta_{12}^{2}-0.04579\,\theta_{34}^{3}
+0.01487\,\theta_{34}^{3}\,\theta_{12}-0.0001593\,\theta_{34}^{4}\nonumber 
\end{eqnarray}

\begin{figure}
\includegraphics[scale=0.7]{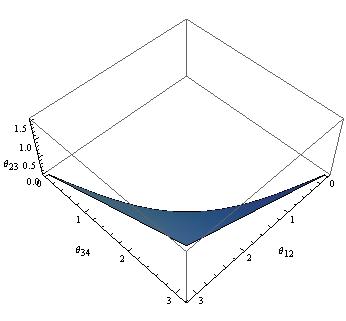}\includegraphics[scale=0.7]{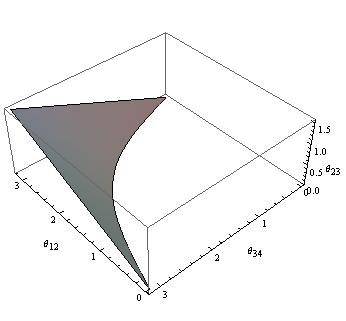}
\caption{Two additional orientations of the Fundamental Surface of Quads.}
\label{figFPQ}
\end{figure}

Figure~\ref{figFPQparam}(b) shows the fit as the gray semi-transparent surface.
The red points are SISell quads corresponding to a random distribution of source positions
within the diamond caustic, on the source plane. As depicted in the Figure, a random 
distribution of source positions does not imply a random distribution of quads on the 
Fundamental Surface; quad density increases with increasing $\theta_{23}$. Two other
orientations of the Fundamental Surface are shown in Figure~\ref{figFPQ}. 

\begin{figure}
\epsscale{0.95}
\plotone{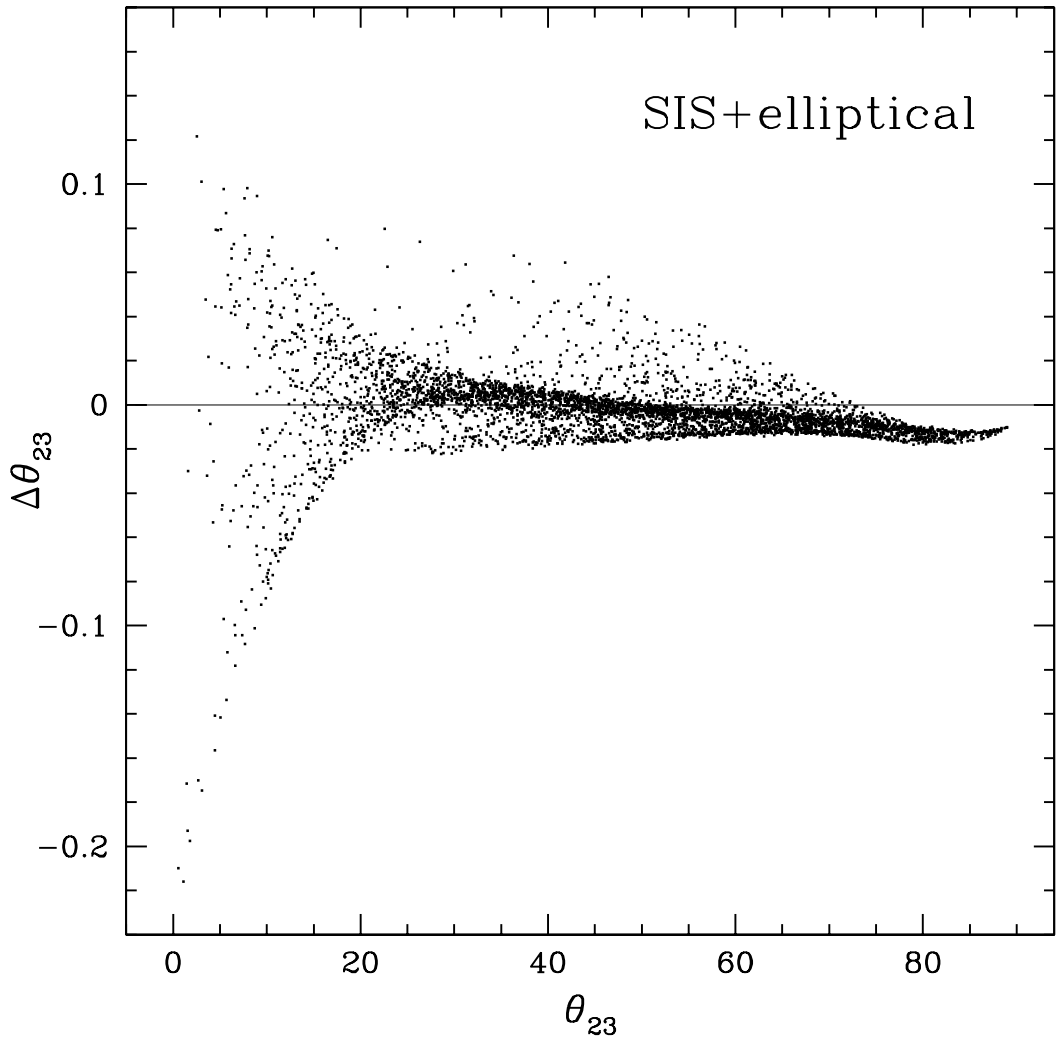}
\vskip-1.3in
\caption{Deviations of the SISell quads from the best fit 4th degree polynomial presented 
in equation~\ref{eq:explicitFunction}. The vertical axis shows the deviations of the 
quads' $\theta_{23}$ from the prediction of equation~\ref{eq:explicitFunction}.
The quads used to compute the best fit were obtained 
using analytical equations for angles $\theta_1$, $\theta_2$, $\theta_3$ and $\theta_4$ 
presented in Section~\ref{sisell}, while the quads in this Figure were generated using a 
ray tracing code. The difference between the two is small.}
\label{theta23Sf}
\end{figure}

Because the RMSE of the SISell quad distribution about the fit plane is $\simlt 0.02^\circ$
the differences between the two will be invisible in the full three dimensional angles space. 
Instead, we calculate the difference in $\theta_{23}$ of the SISell quads and the fitted 
surface keeping the other two angles fixed; we call this difference 
$\Delta\theta_{23}=\theta_{23}-\theta_{23,fit}$, where  $\theta_{23,fit}$ is obtained by 
plugging $\theta_{12}$ and $\theta_{34}$ of a quad in to eq.~\ref{eq:explicitFunction}.
Figure~\ref{theta23Sf} plots $\Delta\theta_{23}$ vs. $\theta_{23}$. The straight horizontal 
line represents the surface fit, eq.~\ref{eq:explicitFunction}, while the points are the 
SISell quads. The wiggles in the distribution of points represents the wiggles in the SISell 
surface, compared to the fourth order polynomial fit. However, for all practical purposes, 
the differences are negligible, and eq.~\ref{eq:explicitFunction} can be taken to be
a good representation of the SISell potential.

We call this surface the Fundamental Surface of Quads because, as we show in the next 
section, not just SISell, but most twofold symmetric models do not differ from it by 
more than a few degrees. This near invariance probably stems from the shape of the
caustic of this class of potentials. The twofold symmetry of the lensing mass
distribution implies the twofold symmetry of the diamond caustic. More specifically
the diagonals of the caustic intersect at $90^\circ$ and all four quadrants of the
caustic are identical.

\section{Other two-fold symmetric potentials}\label{other}

In this Section we explore a wider range of simple twofold symmetric mass models, motivated 
by the commonly used parametric models. We calculate typical values for a total of 12 models,
but show plots (see below) for only eight of these. 

Two of the first four models have isothermal (SIS) radial density profiles, and the other two,
de Vaucouleurs (deV) profile. Isothermal means that if mass ellipticity were zero, the projected 
density profile would scale as $1/r$. The de Vaucouleurs profile has projected density given by,
\begin{equation}
\Sigma=\Sigma_{e}\exp\left(-7.673\left[\left(r/r_{e}\right)^{1/4}-1\right]\right),
\end{equation}
where $r_{e}$ is the half-mass radius, and $\Sigma_{e}$ is the projected mass density 
at $r_{e}$. To generate a diamond caustic the density profiles must be accompanied by 
either ellipticity, $e$, or external shear, $\gamma$. Ellipticity, $e$ of the mass isodensity 
contours is related to the axis ratio of the isodensity contours, $b/a=(1-e)/(1+e)$. 
The properties of the first four models are: 
SIS with $e=0$ and $\gamma=0.3$; 
deV with $e=0$ and $\gamma=0.4$; 
SIE with $e=0.3$ and $\gamma=0$; 
deV with $e=0.4$ and $\gamma=0$.

The additional eight models explore a range of power-law surface mass density profiles, 
$\Sigma(R)\propto R^{-\alpha}$, with $\alpha=0.4,0.7,1.3,1.6$, where isothermal slope is 
$\alpha=1$. The range of slopes we have chosen is considerably wider than what real 
lensing galaxies seem to have. Based on a well defined sample of 15 elliptical galaxy 
lenses, \cite{slacsIII} find that the typical total (dark matter and baryons) space density 
slope is $2.01$, or about $1$ is projection, with dispersion of $0.12$. We chose a range 
of slopes 5 times wider than that because we would like to demonstrate the robustness of 
the FSQ to changes in the lens model parameters. Each of these four density profile slope 
models were given $e=0.25$, or $\gamma=0.25$. These values are somewhat on the high 
side of the typical values for ellipticity and shear encountered in modeling observed quads.

Each mass model generates a two dimensional surface in the three dimensional space of relative 
angles. All surfaces 
coincide at the apex, where $(\theta_{12},\theta_{34},\theta_{23})=(180^\circ,180^\circ,90^\circ)$,
since all twofold symmetric models can produce a perfect cross configuration (though the
distance ratios of images will differ). At other locations the surfaces deviate from each other 
somewhat, and from the one defined by SISell; largest deviations are seen at small
$\theta_{23}$ values, i.e. towards  the base of the triangular surface. However, even at 
the widest separation, the surfaces differ by only a few degrees. The left panel of 
Figure~\ref{type1} shows that these differences are hard to discern in the 
full three dimensional angles space, even if the surface is split up into four pieces for easier
visualization. To make the deviations visible, in the right panel of Figure~\ref{type1}
we fold the surface along the vertical mid-line, and show only a narrow range of angles, 
a few degrees in each case.  In this zoom, the deviations are seen to be a few degrees.

Another way to show deviations is to use $\Delta\theta_{23}$ introduced in the previous Section;
see Figures~\ref{theta23panels} and \ref{theta23panels2}. As before the horizontal line at 
$\Delta\theta_{23}=0$ represents the Fundamental Surface of Quads, given by the fit 
eq.~\ref{eq:explicitFunction}, and the points are quads from lens mass models. 
In Figure~\ref{theta23panels} the isothermal models and the de Vaucouleurs models are 
shown in the left and right panels, respectively. The top row shows the models with external 
shear, $\gamma$, while the bottom rows represent elliptical mass distributions. In
Figure~\ref{theta23panels2} four mass models with power-law density  profiles are shown.

To quantify the effect of shear or ellipticity on any given mass model, we quote the average 
distance ratio $<d_4/d_1>$, where $d_1$ and $d_4$ are the distances of the first and fourth 
arriving images from the lens center, and the average is over quads randomly populating the
inside of the diamond caustic. A given lens mass distribution produces quads with a range of 
image distance ratios, but in general image 1 tends to be farthest from the lens center,
while image 4 tends to be closest. For the 12 models, $<d_4/d_1>$ is between 0.5 and 0.9. 
For the sample of 40 observed quads (Section~\ref{realq}) $<d_4/d_1>\sim 0.68$. i.e. typically
smaller than in our models. As will be discussed in the next Section many of the observed 
quads are the result of mass distributions that are more involved than two-fold symmetric
lenses; most require an external shear in addition to and elliptical lens, while some
require substructure. The presence of these would tend to reduce the $d_4/d_1$ ratio.

Table~\ref{table1} summarizes the results of the 12 models. In general, larger ellipticities
or larger $\gamma$ result in larger deviations from the FSQ.  Because the maximum deviations 
from FSQ differ between models from a fraction of a degree to a few degrees, the range on 
the vertical axes range are different in the top and bottom panels of Figures~\ref{theta23panels} 
and \ref{theta23panels2}. As opposed to the quad distributions generated by the elliptical 
mass models, the ones from models with external shear are much closer to the FSQ, and appear 
more similar to that of the SISell model (unless the surface mass density is very shallow, as in
the bottom right panel of Figure~\ref{theta23panels2}). When viewed in 3D, the surfaces containing 
quads from the elliptical lens models sag below the FSQ, but even for the bottom right panel of 
Figure~\ref{theta23panels} the deviations in $\theta_{23}$ are $<4^\circ$, for $e=0.4$, and
de Vaucoulers profile. Here the images are formed where the projected density $\Sigma(R)\propto R^{-2}$,
or $\rho(r)\propto r^{-3}$ in three dimensions. Given observed lenses, this is a rather extreme 
combination of ellipticity and $\Sigma(R)$. 
Ellipticity of $e=0.4$ corresponds to the axis ratio $b/a=0.43$, or an E5.7 if it were an optical 
elliptical galaxy. Steep density profiles also appear to result in larger deviations from the FSQ;
see Table~\ref{table1}. Real galaxy lenses rarely, if at all, have such steep profiles at the 
location of quad images. For $e=0.25$, and $\Sigma(R)\propto R^{-0.4...-1.6}$ (top right and bottom 
left panels of Figure~\ref{theta23panels2}), the deviations from FSQ are a degree at most.
 
We note that for very large ellipticities or shears (not considered here) the two opposite 
cusps of the diamond caustic protrude
outside of the oval caustic producing so-called naked cusps, which do not produce quads.
The corresponding surfaces of relative angles look similar to the ones without the naked cusps,
except that the portions at the bottom corners of the surface are devoid of quads.

\section{Observed Quads}\label{realq}

In this Section we illustrate one of the practical uses of the Fundamental Surface of Quads
(FSQ).

Galaxy lens systems can be approximately divided into three categories, depending on
whether the lens mass model is (a) an elliptical mass distribution or a circularly 
symmetric mass distribution with an external shear, (b) an elliptical mass distribution 
plus some external shear, or (c) a more complicated mass distribution, possibly with 
additional lens galaxies or substructure mass clumps. A survey of the literature 
indicates that only a handful of systems belong to (a). A model-free way to come to 
that conclusion is to look at the quads in the 3D angles space. 

We have assembled a sample of 40 galaxy-lens quads. The sample was collected from all
available sources, and is therefore heterogeneous.
Where possible, the astrometry, including the positional uncertainties on the images and 
the lensing galaxy was taken from the CASTLeS web-site \citep{Castles}; otherwise from 
individual papers. In the latter case, systems listed in Table~\ref{table2} have a footnote
with a reference; a lens system with no reference means that its data were obtained entirely 
from CASTLeS. The image arrival time was determined from the morphology of the lens \citep{sw03},
and the relative angles were calculated. These are listed in 
Table~\ref{table2}, in columns labeled $\theta_{12}$, $\theta_{34}$ and $\theta_{23}$. 

In Figure~\ref{FPdata} we show two orientations of the 3D angles space with the FSQ
and the 40 quads. For clarity, we do not show errors in this plot. The main conclusion is 
that most observed quads lie more than a few degrees away from the FSQ;
12 are within $\pm 2^\circ$, so most cannot be modeled adequately with an elliptical lens, 
or a circularly symmetric lenses with external shear.

Next, we incorporate errors into the analysis. Even though the astrometric measurement errors 
of images and galaxy lens center are largely independent of each other, a shift in the lens 
center translates into correlated relative angle errors. To account for this we calculate the 
errors as follows. We assume that the $x,~y$ positional errors of each of the four images and
the lens center are normally distributed, with $\sigma_x$ and $\sigma_y$ taken from the 
literature. We then draw thousands of independent image and lens center positions, and for 
each generated lens system calculate relative angles $\theta_{12}$, $\theta_{34}$ and 
$\theta_{23}$. The thousands of generated quads per lens system then give us the error 
distribution for each of the three relative angles. We calculate the mean and the rms 
of these distributions, and list them in columns labeled $\theta_{ij,errors}$ in 
Table~\ref{table2}. Note that the average of these distributions need not be the same as 
$\theta_{ij}$, however, the differences tend  to be small, generally $<0.1^\circ$. 

We quantify the deviation of the quads from the FSQ as in earlier Sections,
by calculating $\Delta\theta_{23}$. The error in 
$\Delta\theta_{23}$, listed as $\Delta\theta_{23,errors}$  is calculated similarly to what 
was described above, using thousands of quads generated based on astrometric uncertainty.

Figure~\ref{theta23data} shows the distribution of the 40 quads in the $\theta_{23}$ 
vs. $\Delta\theta_{23}$ plane. Within errorbars, only 10 systems are consistent with FSQ.
Several of these have published parametric modeling, and are, in fact, well represented by 
two-fold symmetric mass distributions. For example, B2045+265 is successfully modeled by 
\cite{fass99} using SISell potential, eq.~\ref{SISellpot}. SDSS J002240 is modeled by 
\cite{allam07} with Singular Isothermal Ellipsoid (SIE) using {\tt gravlens} software of \cite{k01}.
The same software was used by \cite{grillo10} to fit the positions (not the flux ratios) 
of SDSS J1538 with three types of twofold symmetric models: de Vaucoulers, SIE and a
power law density profile.

On the other hand, some of the lenses which lie away from the FSQ are known to require
external shear in addition to elliptical lens. PG 1115 where the lensing galaxy is a member 
of a galaxy group is inconsistent with FSQ; its $|\Delta\theta_{23}|\sim 4^\circ$.
RXJ 0911+0551 has a cluster next to it \citep{burud98}, so the lens model requires an external 
shear in addition to an elliptical galaxy lens; its $|\Delta\theta_{23}|\sim 6^\circ$. 
For LSD Q0047-2808, \cite{kt03} state that SIE+shear does not fit the image positions well 
(but sufficient for the determination of the Einstein ring radius); it has
$|\Delta\theta_{23}|\sim 14^\circ$.
Lenses with known secondary galaxies also lie far from the FSQ. 
HE 0230-2130 has a secondary lensing galaxy \citep{wis99} very close to the images, and so 
its $|\Delta\theta_{23}|\sim 30^\circ$. B1608 has a complicated galaxy merger as a lens,
and its $|\Delta\theta_{23}|\sim 5^\circ$, and so it is inconsistent with the FSQ. 

A few caveats are in order. 
If a quad does not lie within a couple of degrees of the Fundamental Surface of Quads 
(i.e. the range defined by the various density profile and ellipticity models, such as the ones
in Figure~\ref{type1}, \ref{theta23panels} and \ref{theta23panels2}), 
it cannot be modeled by a twofold symmetric lens. However, the opposite 
need not be true. If a quad lies on the FSQ does that immediately imply
that it can be modeled by a twofold symmetric lens, regardless of its image distance ratios?
This question will be addressed in a future study. We also note that even if 
a quad does belong to the locus of twofold symmetric lens in the full 6D space of
image position properties, it does not mean that other types of lens models cannot
fit it. Reconstruction of the lens mass reconstruction from single quads is a highly 
underconstrained problem, so the solution is not unique, and many mass models can reproduce 
the image positions exactly \citep{sw04}.

\section{Conclusions and Future Work}

In this paper we present a model-free way of making inferences about the lensing mass 
distribution given its quad image positions. The latter are represented by three
relative angles that describe the distribution of images around the lens center. We show 
that in the three dimensional space of these angles, quads generated by SIS+elliptical
mass distribution belong to an invariant two dimensional surface, regardless of the
shear parameter $\gamma$, and normalization $b$. Furthermore, quads from a wider class 
of lenses with twofold symmetry outline almost the same surface, making the surface a 
near invariant descriptor of twofold symmetric mass distributions. Because of that 
property we call the two dimensional surface the Fundamental Surface of Quads (FSQ). 

\begin{figure}
\includegraphics[scale=0.50]{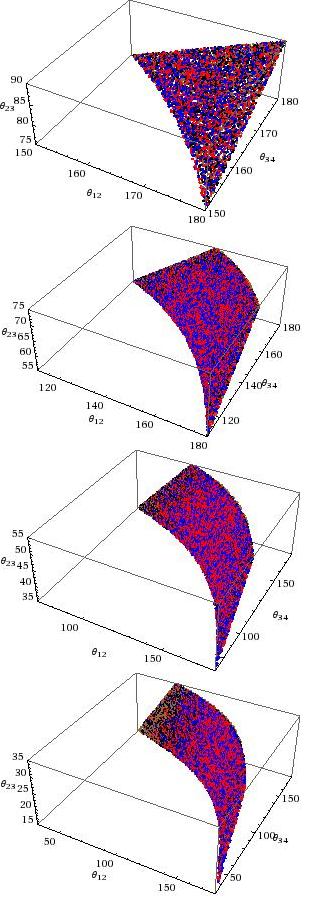}\includegraphics[scale=0.50]{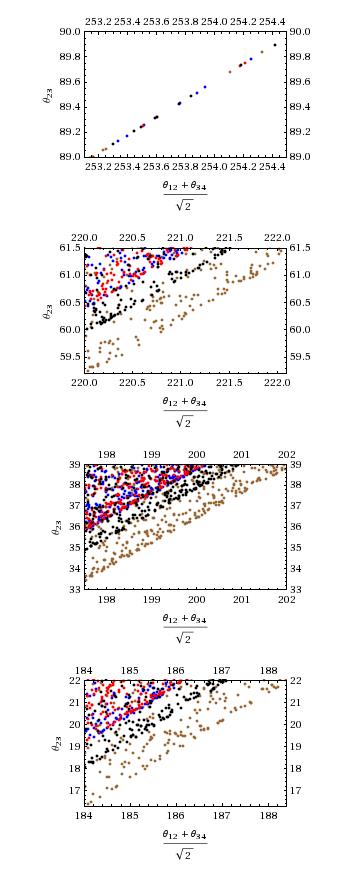}
\caption{Simple twofold symmetric lens mass distributions define a nearly
invariant Fundamental Surface of Quads. Quads from four mass models are shown:
SIS with $e=0$ and $\gamma=0.3$ (red); 
deV with $e=0$ and $\gamma=0.4$ (blue); 
SIE with $e=0.3$ and $\gamma=0$ (black);
deV with $e=0.4$ and $\gamma=0$ (brown).
On the left we show the 3D space of relative angles sliced
into four segments divided by $\theta_{23}=35^\circ ,55^\circ, 75^\circ$.
The fact that the points of different lens potentials are hard to
tell apart demonstrates the near invariance of the FSQ.
To make the deviations visible, on the right we fold and project a small angle 
range of the surface; see Section~\ref{other} for details.}
\label{type1}
\end{figure}

\begin{figure}
\epsscale{0.9}
\plotone{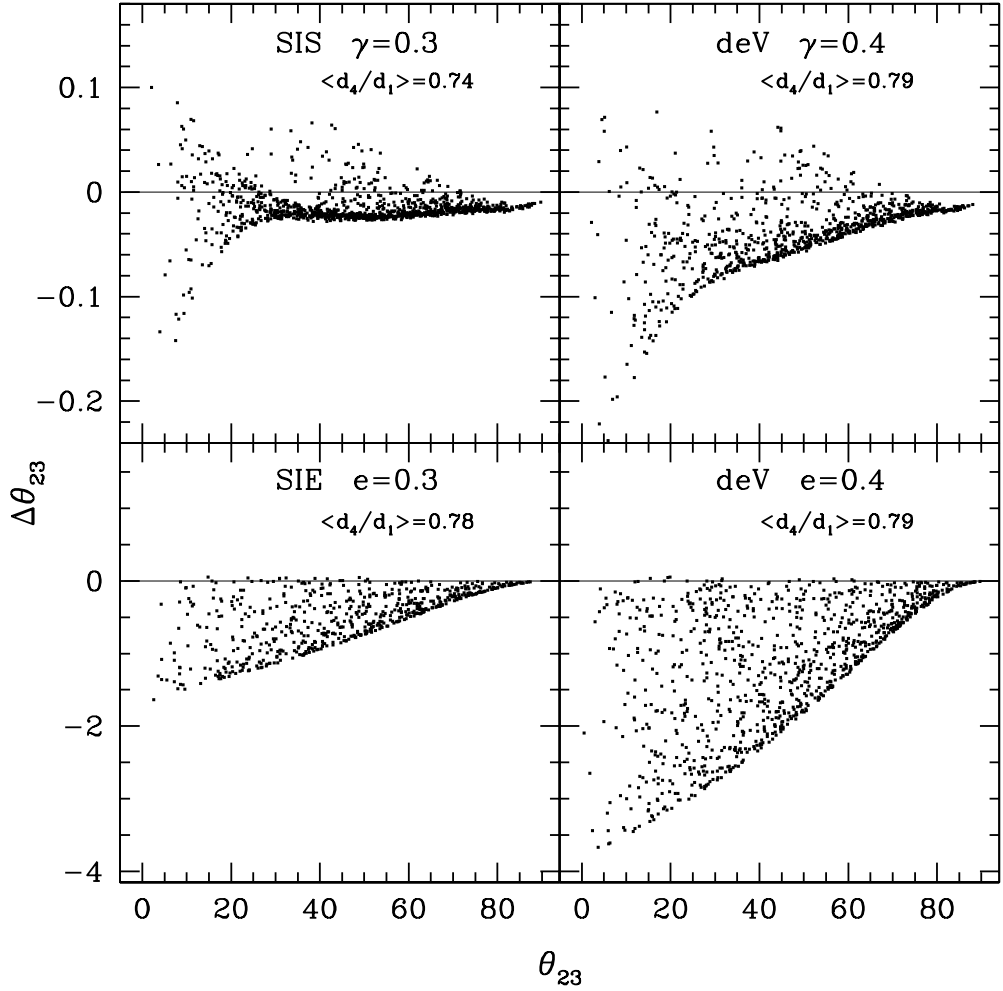}
\vskip-1.3in
\caption{The deviations of the quads of four mass distributions from the predictions
of the 4th degree polynomial fit, equation~\ref{eq:explicitFunction}. The four 
mass distributions are the same as the ones shown in Figure~\ref{type1}
{\it Top left:} Circularly symmetric SIS surface mass density with external shear $\gamma=0.3$;
{\it Top right:} Circularly symmetric de Vaucoulers with external shear $\gamma=0.4$;
{\it Bottom left:} Elliptical SIE with ellipticity 0.3;
{\it Bottom left:} Elliptical de Vaucoulers with ellipticity 0.4.
The average value of the distance ratio of the fourth to first arriving image, $<d_4/d_1>$,
is shown in each panel. This aids in visualizing the meaning of $\gamma$ and $e$ value.
Note that the vertical axes have different ranges in the top and bottom panels.}
\label{theta23panels}
\end{figure}

\begin{figure}
\epsscale{0.9}
\plotone{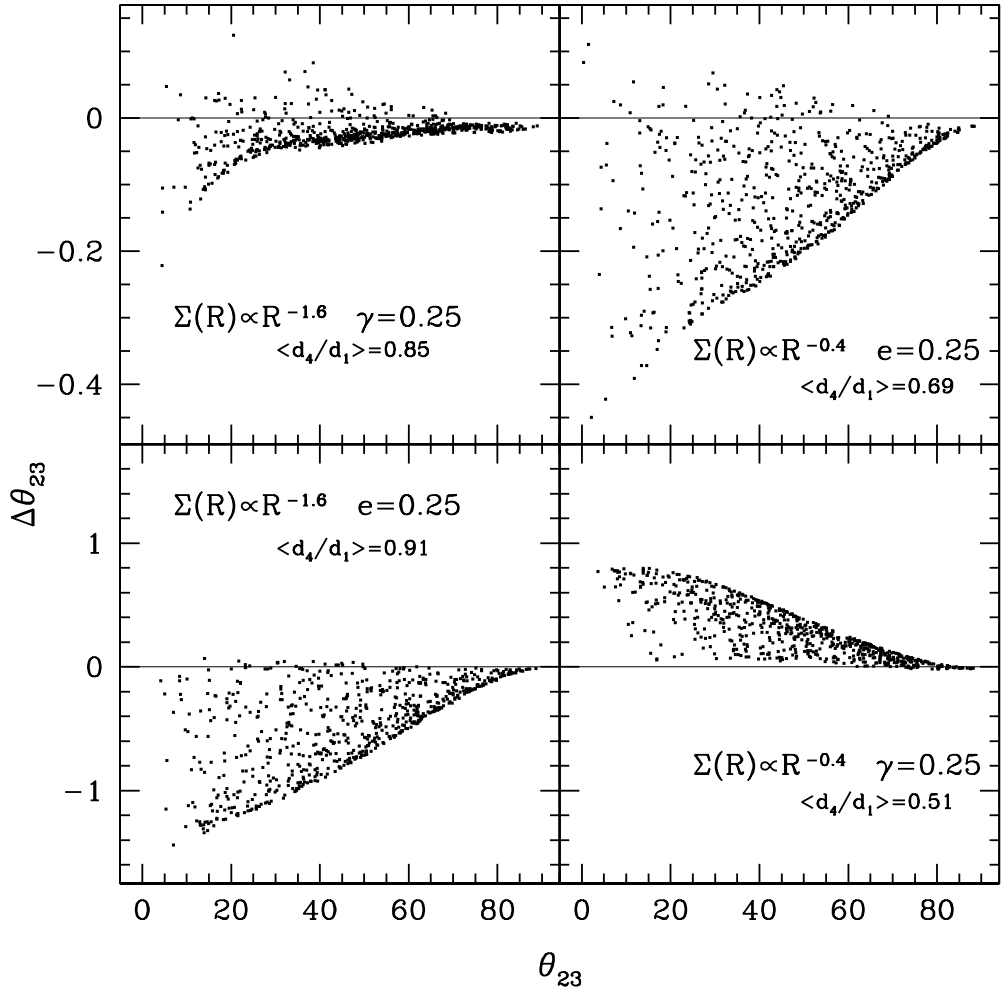}
\vskip-1.3in
\caption{Same as Figure~\ref{theta23panels}, but for a different set of lens mass models.
{\it Top left:} Circularly symmetric surface mass density, $\Sigma(R)\propto R^{\,-1.6}$ with external shear $\gamma=0.25$;
{\it Top right:} Elliptical $\Sigma(R)\propto R^{\,-0.4}$ with ellipticity 0.25;
{\it Bottom left:} Elliptical $\Sigma(R)\propto R^{\,-1.6}$ with ellipticity 0.25;
{\it Bottom left:} Circularly symmetric $\Sigma(R)\propto R^{\,-0.4}$ with external shear $\gamma=0.25$.}
\label{theta23panels2}
\end{figure}

\begin{figure}
\includegraphics[scale=0.6]{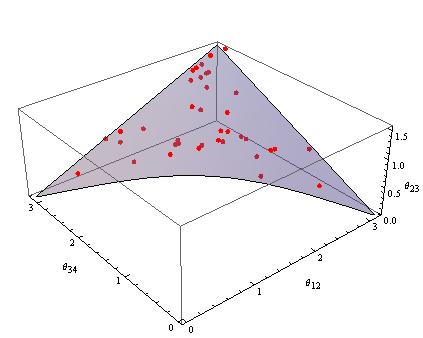}\includegraphics[scale=0.6]{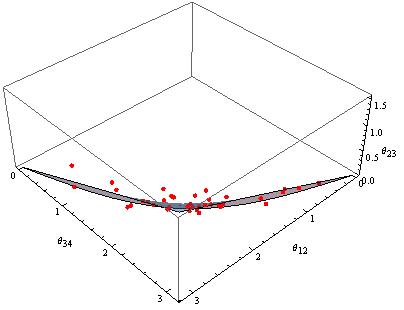}
\caption{The Fundamental Surface of Quads fit equation (shaded region) and the forty observed 
galaxy quads (red dots). Two orientations are shown; in the second one the deviation
of the observed quads from the Plane are clearly visible.}
\label{FPdata}
\end{figure}

\begin{figure}
\epsscale{0.95}
\plotone{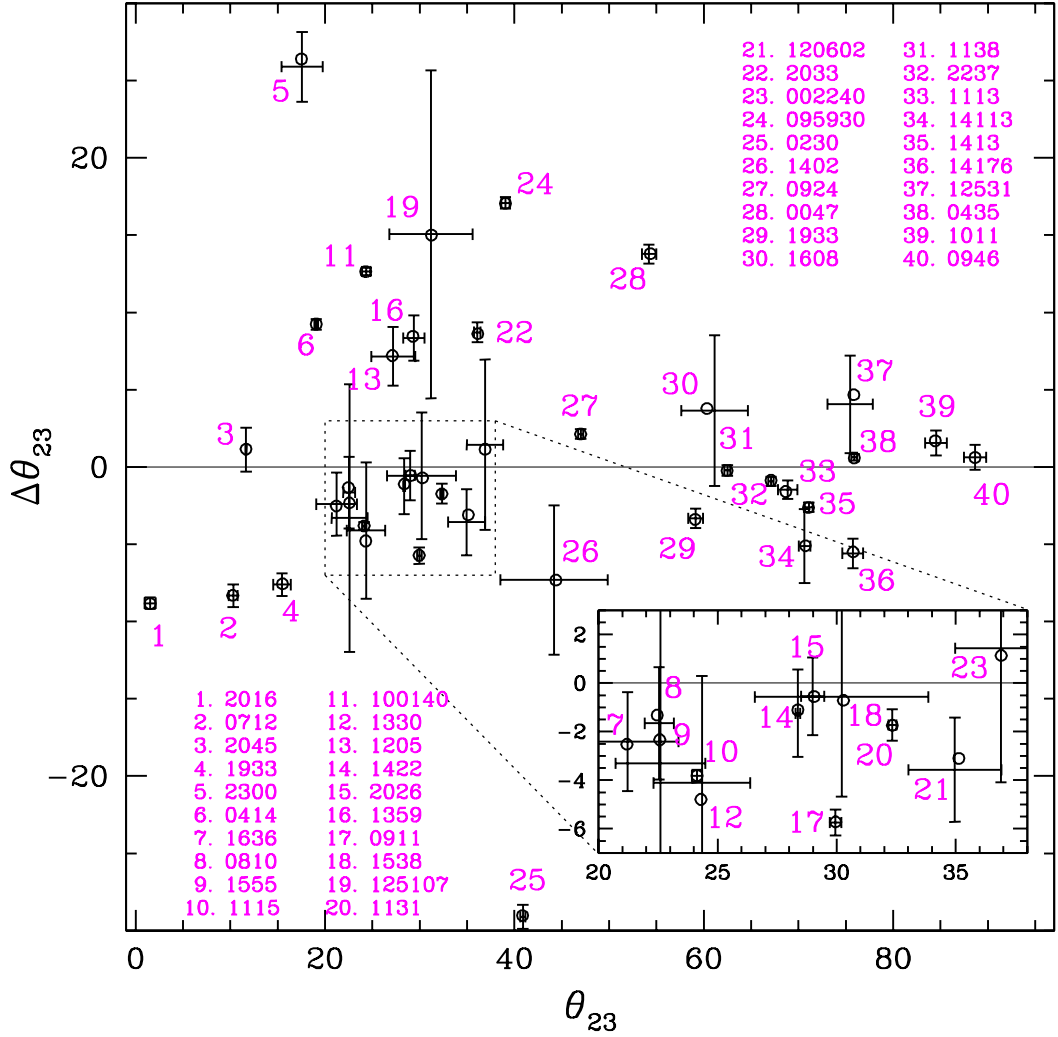}
\vskip-1.3in
\caption{Deviations, $\Delta\theta_{23}$, of the observed quads from the FSQ. 
The horizontal axis is the observed $\theta_{23}$. The empty circles represent the observed
relative angles and their deviations from the FSQ. The error bars are calculated as explained 
in Section~\ref{realq}. The horizontal and vertical axes values are given in Table~\ref{table1}.}
\label{theta23data}
\end{figure}

The existence of FSQ allows one to characterize galaxies and clusters based on the 
quads they generate. To aid in that, we provide a fitting formula for the FSQ based 
on the SIS+elliptical lensing potential. If a quad does not lie within a couple of degrees 
of the FSQ (i.e. the range defined by the various density profile and ellipticity models, 
such as the ones in Figure~\ref{type1}, \ref{theta23panels}, and \ref{theta23panels2}), the mass
distribution is not twofold symmetric. This method of determining if a lens can be fit 
with a twofold symmetric lens is superior to answering this question using parametric 
modeling. The latter fits quads with a finite set of models, while our method addresses 
all twofold symmetric models irrespective of the specific parametric form.

However, the main importance of the FSQ is not in ascertaining if the lens mass 
distribution is twofold symmetric or not, but in the following aspects, which we 
will investigate in the forthcoming papers.
First, the near invariance of FSQ provides a new framework for studying quads, and strong 
lensing theory in general. We remind the reader that it is still not understood why a wide class
of twofold symmetric lenses form such a tight, near invariant distribution in the space of 
relative angles. 
Second, as already shown in \cite{wffb08}, the relative angles present a promising way of 
investigating realistic mass distributions, and specifically, differentiating substructured
lenses from smooth non-twofold symmetric ones. 
Finally,  the full set of quad image properties lives in the six dimensional space that
includes image distance ratios. An investigation of this space is yet to be undertaken.

\clearpage{}

\begin{deluxetable}{cccccc}
\rotate
\tablewidth{0pc}
\tablecaption{Summary of lens mass models}
\tablehead{
\colhead {Lens Mass Model} & $<d_4/d_1>$ & rms $(0^\circ\!\!-\!90^\circ)$ & rms $(0^\circ\!\!-\!30^\circ)$ & rms $(30^\circ\!\!-\!60^\circ$)& rms $(60^\circ\!\!-\!90^\circ)$}
\startdata
SIS with $\gamma=0.3$ ($\Sigma\propto R^{-1}$)  & 0.74 & 0.025 & 0.042 & 0.021  & 0.016 \\  
SIE with $e=0.3$      ($\Sigma\propto R^{-1}$)  & 0.78 & 0.601 & 0.877 & 0.600  & 0.223 \\  
deV with $\gamma=0.4$ ($\Sigma\propto R^{-3}$)  & 0.79 & 0.0526 & 0.0849 & 0.0487 & 0.0224 \\  
deV with $e=0.4$      ($\Sigma\propto R^{-2}$)  & 0.79 & 1.431 & 1.948 & 1.457  & 0.541 \\  
$\Sigma(R)\propto R^{\,-0.4}$ with $e=0.25$     & 0.69 & 0.148 & 0.211 & 0.153  & 0.0676 \\  
$\Sigma(R)\propto R^{\,-0.7}$ with $e=0.25$     & 0.76 & 0.322 & 0.475 & 0.323  & 0.125 \\  
$\Sigma(R)\propto R^{\,-1.3}$ with $e=0.25$     & 0.91 & 0.547 & 0.802 & 0.548 & 0.210 \\  
$\Sigma(R)\propto R^{\,-1.6}$ with $e=0.25$     & 0.91 & 0.565 & 0.810 & 0.564 & 0.222 \\  
$\Sigma(R)\propto R^{\,-0.4}$ with $\gamma=0.25$& 0.51 & 0.356 & 0.603 & 0.347 & 0.100 \\  
$\Sigma(R)\propto R^{\,-0.7}$ with $\gamma=0.25$& 0.85 & 0.0432 & 0.0716 & 0.0403 & 0.0101 \\  
$\Sigma(R)\propto R^{\,-1.3}$ with $\gamma=0.25$& 0.82 & 0.0592 & 0.0816 & 0.0583 & 0.413 \\  
$\Sigma(R)\propto R^{\,-1.6}$ with $\gamma=0.25$& 0.85 & 0.0342 & 0.0591 & 0.0282 & 0.0159 \\  
\enddata 
\vskip0.2in
{The first four lens models are shown in Figure~\ref{theta23panels}, and a subset of the other eight are in Figure~\ref{theta23panels2}. For the first four the slope of the projected surface mass density, $\Sigma(R)$, at the location of the images is indicated in parentheses. For de Vaucouleurs models the slope of the density profile changes with radius, so the value of the slope is a typical value for the radii where the images form. The other eight density profiles are power-laws in radius. The quantity $<d_4/d_1>$ is the average ratio of the distance of the 4th and 1st arriving images. In the third column, rms $(0^\circ\!\!-\!90^\circ)$ is the rms value of $\Delta\theta_{23}$ for all the images, i.e. those with the full range of $\theta_{23}$, or $0^\circ\!\!-\!90^\circ$. The last three columns show rms values for three subsets of images, divided by their $\theta_{23}$ values. All rms are quoted in degrees.}
\label{table1}
\end{deluxetable}

\clearpage

\begin{deluxetable}{cccccccccc}
\rotate
\tablewidth{0pc}
\tablecaption{Relative angles}
\tablehead{
\colhead 
 N & {Lens name} & $\theta_{12}$ & $\theta_{12,{\rm{errors}}}$ & $\theta_{34}$ & $\theta_{34,{\rm{errors}}}$ & $\theta_{23}$ & $\theta_{23,{\rm{errors}}}$ & $\Delta\theta_{23}$ & $\Delta\theta_{23,{\rm{errors}}}$}
\startdata
 1 & MG 2016+112     & 150.8 & 150.7 $\pm$ 0.4  &  91.4 &  91.3 $\pm$ 0.6  &  1.5 &   1.5 $\pm$ 0.6  & -8.8 &  -8.8 $\pm$ 0.4\\
 2 & B 0712+472      &  79.8 &  79.8 $\pm$ 0.3  & 163.2 & 163.3 $\pm$ 0.7  & 10.2 &  10.3 $\pm$ 0.4  & -8.3 &  -8.3 $\pm$ 0.7\\
 3 & B 2045+265      &  34.9 &  34.9 $\pm$ 0.1  & 175.2 & 175.2 $\pm$ 1.0  & 11.7 &  11.7 $\pm$ 0.1  &  1.2 &   1.1 $\pm$ 1.4\\
 4 & B 1933+503 lobe & 155.5 & 155.5 $\pm$ 0.7  & 101.7 & 101.7 $\pm$ 0.9  & 15.5 &  15.5 $\pm$ 0.9  & -7.6 &  -7.6 $\pm$ 0.7\\
 5 & SLACS J2300+002 & 160.8 & 161.2 $\pm$ 2.0  &  38.5 &  38.5 $\pm$ 2.0  & 17.5 &  17.6 $\pm$ 2.2  & 26.4 &  25.9 $\pm$ 2.3\\
 6 & MG 0414+0534    & 101.5 & 101.5 $\pm$ 0.3  & 144.1 & 144.1 $\pm$ 0.3  & 19.1 &  19.1 $\pm$ 0.2  &  9.2 &   9.2 $\pm$ 0.3\\
 7 & SLACS J1636+470 & 128.0 & 127.9 $\pm$ 1.8  & 136.9 & 136.9 $\pm$ 2.5  & 21.2 &  21.2 $\pm$ 2.1  & -2.5 &  -2.4 $\pm$ 2.0\\
 8 & HS 0810+2554    & 111.3 & 111.5 $\pm$ 1.1  & 150.1 & 150.4 $\pm$ 2.3  & 22.5 &  22.5 $\pm$ 0.6  & -1.3 &  -1.7 $\pm$ 2.3\\
 9 & B 1555+375      & 114.0 & 113.1 $\pm$ 5.9  & 149.3 & 150.6 $\pm$10.7  & 22.6 &  22.6 $\pm$ 1.9  & -2.3 &  -3.3 $\pm$ 8.7\\
10 & PG 1115+080     & 141.9 & 141.9 $\pm$ 0.3  & 127.5 & 127.5 $\pm$ 0.4  & 24.1 &  24.1 $\pm$ 0.2  & -3.8 &  -3.8 $\pm$ 0.3\\
11 & J 100140.12+020 & 120.4 & 120.4 $\pm$ 0.1  & 131.3 & 131.3 $\pm$ 0.5  & 24.3 &  24.3 $\pm$ 0.4  & 12.6 &  12.7 $\pm$ 0.2\\
12 & SDSS J1330+1810 & 115.6 & 115.7 $\pm$ 3.2  & 152.2 & 151.5 $\pm$ 4.8  & 24.3 &  24.3 $\pm$ 2.0  & -4.8 &  -4.1 $\pm$ 4.4\\
13 & SLACS J1205+491 & 159.1 & 159.3 $\pm$ 1.9  &  90.4 &  90.3 $\pm$ 2.1  & 27.1 &  27.2 $\pm$ 2.3  &  7.2 &   7.2 $\pm$ 1.9\\
14 & B 1422+231      &  74.8 &  74.8 $\pm$ 0.3  & 173.8 & 174.0 $\pm$ 1.4  & 28.4 &  28.4 $\pm$ 0.1  & -1.1 &  -1.2 $\pm$ 1.8\\
15 & WFI 2026-4536   & 154.1 & 154.1 $\pm$ 1.8  & 113.6 & 113.5 $\pm$ 1.3  & 29.1 &  29.0 $\pm$ 0.5  & -0.5 &  -0.5 $\pm$ 1.6\\
16 & CLASS B1359+154 & 135.9 & 135.8 $\pm$ 1.2  & 125.8 & 126.1 $\pm$ 3.2  & 29.3 &  29.4 $\pm$ 1.1  &  8.4 &   8.3 $\pm$ 1.5\\
17 & RXJ 0911+0551   & 180.7 & 180.7 $\pm$ 0.5  &  69.6 &  69.7 $\pm$ 0.4  & 29.9 &  29.9 $\pm$ 0.2  & -5.7 &  -5.7 $\pm$ 0.5\\
18 & SDSS J1538+5817 & 152.7 & 152.7 $\pm$ 4.0  & 117.7 & 117.3 $\pm$ 4.0  & 30.3 &  30.2 $\pm$ 3.6  & -0.7 &  -0.6 $\pm$ 4.1\\
19 & SDSS J125107    & 158.8 & 158.1 $\pm$ 9.9  &  85.2 &  86.4 $\pm$ 8.8  & 31.3 &  31.2 $\pm$ 4.4  & 15.0 &  15.1 $\pm$10.6\\
20 & RXJ 1131-1231   &  66.0 &  66.0 $\pm$ 0.2  & 180.9 & 180.8 $\pm$ 0.5  & 32.3 &  32.3 $\pm$ 0.2  & -1.8 &  -1.7 $\pm$ 0.6\\
21 & SDSS J120602.09 &  96.0 &  95.9 $\pm$ 2.2  & 171.6 & 171.9 $\pm$ 2.1  & 35.1 &  35.0 $\pm$ 2.0  & -3.1 &  -3.6 $\pm$ 2.1\\
22 & WFI 2033-4723   & 140.6 & 140.5 $\pm$ 0.6  & 128.5 & 128.5 $\pm$ 0.7  & 36.1 &  36.1 $\pm$ 0.3  &  8.6 &   8.7 $\pm$ 0.6\\
23 & SDSS J002240    &  77.9 &  78.3 $\pm$ 1.8  & 177.5 & 177.1 $\pm$ 4.4  & 36.9 &  36.9 $\pm$ 1.9  &  1.1 &   1.4 $\pm$ 5.5\\
24 & J 095930.94+023 & 141.2 & 141.2 $\pm$ 0.4  & 120.9 & 121.0 $\pm$ 0.5  & 39.0 &  39.1 $\pm$ 0.5  & 17.0 &  17.1 $\pm$ 0.4\\
25 & HE 0230-2130    & 127.2 & 127.2 $\pm$ 0.3  & 186.7 & 186.8 $\pm$ 0.9  & 40.9 &  40.9 $\pm$ 0.3  &-29.0 & -29.1 $\pm$ 0.8\\
26 & SDSS 1402+6321  & 142.6 & 142.7 $\pm$ 6.1  & 156.1 & 155.7 $\pm$ 6.8  & 44.4 &  44.2 $\pm$ 5.6  & -7.3 &  -7.3 $\pm$ 4.8\\
27 & SDSS 0924+0219  & 153.6 & 153.6 $\pm$ 0.4  & 135.9 & 135.9 $\pm$ 0.5  & 47.0 &  47.0 $\pm$ 0.3  &  2.1 &   2.1 $\pm$ 0.3\\
28 & LSD Q0047-2808  & 130.9 & 130.8 $\pm$ 0.7  & 152.8 & 152.7 $\pm$ 0.8  & 54.3 &  54.2 $\pm$ 0.8  & 13.8 &  13.8 $\pm$ 0.6\\
29 & B 1933+503 core & 169.1 & 169.1 $\pm$ 0.8  & 142.8 & 142.8 $\pm$ 0.8  & 59.1 &  59.1 $\pm$ 0.8  & -3.4 &  -3.3 $\pm$ 0.6\\
30 & B 1608+656      &  97.9 &  99.2 $\pm$ 5.9  & 186.7 & 187.2 $\pm$ 4.9  & 60.3 &  61.1 $\pm$ 3.5  &  3.8 &   3.7 $\pm$ 4.9\\
31 & SDSS 1138+0314  & 153.1 & 153.1 $\pm$ 0.5  & 161.1 & 161.1 $\pm$ 0.6  & 62.5 &  62.5 $\pm$ 0.3  & -0.2 &  -0.2 $\pm$ 0.4\\
32 & Q 2237+0305     & 146.3 & 146.3 $\pm$ 0.4  & 173.4 & 173.5 $\pm$ 0.6  & 67.1 &  67.1 $\pm$ 0.3  & -0.9 &  -0.9 $\pm$ 0.3\\
33 & HE 1113-0641    & 154.4 & 154.6 $\pm$ 1.2  & 170.3 & 170.3 $\pm$ 1.2  & 68.6 &  68.9 $\pm$ 1.0  & -1.6 &  -1.5 $\pm$ 0.6\\
34 & HST 14113+5211  & 163.2 & 163.2 $\pm$ 0.6  & 171.0 & 171.0 $\pm$ 3.8  & 70.7 &  70.6 $\pm$ 0.6  & -5.1 &  -5.1 $\pm$ 2.4\\
35 & H 1413+117      & 160.3 & 160.3 $\pm$ 0.6  & 170.3 & 170.4 $\pm$ 0.7  & 71.1 &  71.1 $\pm$ 0.4  & -2.6 &  -2.6 $\pm$ 0.3\\
36 & HST 14176+5226  & 163.1 & 163.1 $\pm$ 0.5  & 179.5 & 179.6 $\pm$ 1.6  & 75.8 &  75.7 $\pm$ 1.1  & -5.5 &  -5.6 $\pm$ 1.0\\
37 & HST 12531-2914  & 149.9 & 149.6 $\pm$ 2.2  & 175.0 & 175.5 $\pm$ 4.4  & 75.8 &  75.4 $\pm$ 2.4  &  4.7 &   4.1 $\pm$ 3.2\\
38 & HE 0435-1223    & 155.1 & 155.2 $\pm$ 0.3  & 176.8 & 176.7 $\pm$ 0.3  & 75.9 &  75.9 $\pm$ 0.3  &  0.6 &   0.6 $\pm$ 0.2\\
39 & SDSS 1011+0143  & 169.7 & 169.9 $\pm$ 0.9  & 176.4 & 176.6 $\pm$ 1.5  & 84.4 &  84.5 $\pm$ 1.2  &  1.7 &   1.6 $\pm$ 0.8\\
40 & SLACS J0946+100 & 182.9 & 182.9 $\pm$ 0.9  & 172.8 & 172.7 $\pm$ 1.1  & 88.6 &  88.6 $\pm$ 1.2  &  0.6 &   0.6 $\pm$ 0.8\\
\enddata
\tablerefs{
MG 2016+112 \citep{ng97,law84},
B2045+265 \citep{fass99},
B 1933+503 \citep{n98},
SLACS J2300+002, SLACS J1636+470, SLACS J1205+491 \citep{fsb08},
SDSS J125107 \citep{k07},
SDSS J1330+1810 \citep{ogu08},
J 100140.12+020 \citep{jack08},
SDSS J1538+5817 \citep{grillo10},
SDSS J002240 \citep{allam07},
SDSS J120602.09 \citep{lin09},
J 095930.94+023 \citep{jack08},
SDSS 1402+6321 \citep{bol05},
LSD Q0047-2808 \citep{kt03},
HE 1113-0641  \citep{black08},
SLACS J0946+100 (outer ring) \citep{gav08}
} 
\label{table2}
\end{deluxetable}

\clearpage


\end{document}